\documentclass[twoside,slac_one]{revtex4}
\usepackage{graphicx}
\usepackage{fancyhdr}
\usepackage{amsmath} 
\usepackage{bm}
\usepackage{amsxtra}
\usepackage{amssymb}
\usepackage{amsthm}
\usepackage{latexsym}
\usepackage{lscape}

\def\MET{{\mbox{$E\kern-0.57em\raise0.19ex\hbox{/}_{T}$}}}
\def\METwithSpace{{\mbox{$E\kern-0.57em\raise0.19ex\hbox{/}_{T}~$}}}
\newcommand{\comphep}   {\sc{c}\rm{omp}\sc{hep}}
\newcommand{\ttbar}     {\mbox{$t\bar{t}$}}

\begin{document}

\title{Flavor changing neutral currents in ttbar decays at D\O}

%

\author{C.L. McGivern for the D0 Collaboration}
\affiliation{The University of Kansas, Lawrence, KS 66045 USA}

\begin{abstract}
We present a search for flavor changing neutral currents (FCNC) in decays of top 
quarks. The analysis is based on a search for 
$t\bar{t}\rightarrow\ell'\nu\ell\bar{\ell}$+jets ($\ell, \ell' = e,\mu$) final 
states using 4.1~{\rm fb}$^{-1}$ of integrated luminosity of $p\bar{p}$
collisions at $\sqrt{s} = 1.96$~{\rm TeV}. We extract limits on the branching 
ratio $B(t\rightarrow Zq)$ ($q = u, c$ quarks), assuming anomalous $tuZ$ 
or $tcZ$ couplings. We do not observe any sign of such anomalous coupling 
and set a limit of $B < 3.2\%$ at 95\% C.L.
\end{abstract}

\maketitle

\thispagestyle{fancy}


\section{Introduction}
In this paper, we search for FCNC decays of the top ($t$) quark~\cite{topdecay}. 
Within the standard model (SM) the top quark decays into a $W$ boson and a $b$ 
quark with a rate proportional to the Cabibbo-Kobayashi-Maskawa (CKM) matrix element 
squared, $|V_{\rm tb}|^2$~\cite{topdecay}. Under the assumption of three fermion families 
and a unitary $3 \times 3$ CKM matrix, the $|V_{\rm tb}|$ element is severely 
constrained to $|V_{\rm tb}| = 0.999152^{+0.000030}_{-0.000045}$~\cite{pdg}.
While the SM branching fraction for $t\rightarrow Zq$ ($q = u, c$ quarks) is 
predicted to be $\approx 10^{-14}$~\cite{tZqfcnc}, supersymmetric extensions 
of the SM with or without $R$-parity violation, or quark compositeness predict 
branching fractions as high as $\approx 10^{-4}$~\cite{tZqfcnc,BSMtop1,fcnclhc}. 
The observation of the FCNC decay $t \rightarrow Zq$ would therefore provide 
evidence of contributions from beyond SM (BSM) physics.

We analyze top-pair production ($t\overline{t}$), where either one or both of the 
top quarks decay via $t \rightarrow Zq$ or their charge conjugates (hereafter 
implied). Any top quark that does not decay via $t \rightarrow Zq$ is assumed 
to decay via $t \rightarrow Wb$. We assume that the $t\rightarrow Zq$ decay is 
generated by an anomalous FCNC term added to the SM Lagrangian
\begin{equation}
\label{eq:fcnc_lagrangian}
{\cal L}_{\rm FCNC} = \frac{e}{2 \sin\theta_W \cos\theta_W} \ \bar{t} \, 
   \gamma_\mu ( v_{tqZ} - a_{tqZ} \gamma_5 )\ q \, Z^\mu \,\, + h.c.,
\end{equation}
where $q$, $t$, and $Z$ are the quantum fields for up or charm quarks, top quarks, 
and for the $Z$ boson, respectively, $e$ is the electric charge, and $\theta_W$ the 
Weinberg angle. We thereby introduce dimension-4 vector, $v_{tqZ}$, and axial 
vector, $a_{tqZ}$, couplings as defined in~\cite{fcnc_coupling}. We find 
in Refs.~\cite{nloqcd1, nloqcd2} that the next-to-leading order (NLO) effects due to 
perturbative QCD corrections are negligible when extracting the branching ratio 
limits to the leading order (LO) in Eq.~\ref{eq:fcnc_lagrangian}.

\begin{figure}[ht]
\centering
  \includegraphics[width=0.28\textwidth]{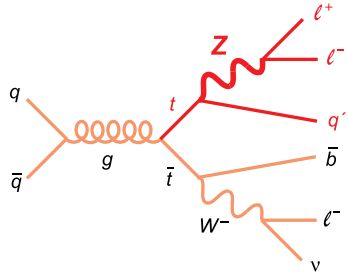}
\caption{Lowest-order diagram for FCNC $t\overline{t} \rightarrow WbZq'$ 
production, where $q'$ can be either a $u$ or $c$ quark, and the $W$ and $Z$ 
bosons decay leptonically.}\label{fig:FCNCFeyn}
\end{figure}

We investigate channels where the $W$ and $Z$ bosons decay leptonically, as 
shown in Fig.~\ref{fig:FCNCFeyn}. The $u$, $c$, and $b$ quarks subsequently 
hadronize, giving rise to a final state with three charged leptons ($\ell = e, \mu$), 
an imbalance in momentum transverse to the $p\bar{p}$ collision axis ($\MET$, 
assumed to be from the escaping neutrino in the $W\rightarrow \ell\nu$ decay), 
and jets. 

This is the first search for FCNC in $t\bar{t}$ decays with trilepton final states. 
This mode provides a distinct signature with low background, albeit at the cost 
of statistical power. The first measurement ($b \rightarrow s\gamma$) was 
published in 1995 by the CLEO Collaboration~\cite{cleo1995}. Numerous studies 
have been done since then to search for FCNC processes in meson decays, i.e., 
$b \rightarrow Zs$ in $B^{+} \rightarrow K^{*+}\ell^{+} \ell^{-}$~\cite{Bmeson_cdf, 
Bmeson_belle, Bmeson_babar}, $B \rightarrow K^{*}\nu\bar{\nu}$~\cite{kamenik_arXiv}, 
and $B_{s,d} \rightarrow \ell^{+} \ell^{-}$~\cite{Bmeson_artuso, Bmeson_antonelli}
or $s \rightarrow Zd$ in $K^{+} \rightarrow \pi^{+} \nu \bar{\nu}$~\cite{Kmeson_e949}. 
Using the $D^{+} \rightarrow \pi^{+} \mu^{+} \mu^{-}$ final state in 
$1.3$~{\rm fb}$^{-1}$ of integrated luminosity, the D0 Collaboration has set the 
best branching ratio ($B$) limits on the FCNC $c\rightarrow Zu$ process at 
$B(D^{+} \rightarrow \pi^{+} \mu^{+} \mu^{-} ) < 3.9 \times 10^{-6}$ at 90\% 
C.L.~\cite{fcnc_Dmeson}. There are theoretical arguments as to why top quark decays 
may be the best way to study flavor violating couplings of mass-dependent 
interactions~\cite{rsmodelitself, fcnc_rsmodel}. FCNC $tqZ$ and $t q \gamma$ 
couplings have been studied by the CERN $e^+e^-$ Collider (LEP), DESY $ep$ 
Collider (HERA), and Fermilab $p\bar{p}$ Collider (Tevatron) experiments~\cite{fcnc_lep,fcnc_h1,fcnc_zeus,fcnc_tqgamma_cdf,cdflimits}.
The D0 Collaboration has recently published limits on the branching ratios 
determined from FCNC gluon-quark couplings using single top quark final 
states~\cite{d0glimits}. The $95\%$ C.L. upper limit on the branching ratio of 
$t \rightarrow Zq$ from the CDF Collaboration uses 1.9~{\rm fb}$^{-1}$ of 
integrated luminosity, assumes a top quark mass of $m_{\rm t} = 175$~{\rm GeV} 
and uses the measured cross section of $\sigma_{t\bar{t}} = 8.8 \pm 1.1$ 
{\rm pb}~\cite{cdflimits}. This result excludes branching ratios of 
$B(t\rightarrow Zq) > 3.7\%$, with an expected limit of 5.0\% $\pm$ 2.2\%. 
To obtain these results, CDF exploited the two lepton plus four jet final state. 
This signature occurs when one of the pair-produced top quarks decays via 
FCNC to $Zq$, followed by the decay $Z \rightarrow ee$ or $Z \rightarrow \mu\mu$. 
The other top quark decays to $Wb$, followed by the hadronic decay of the $W$ 
boson. This dilepton signature suffers from large background, but profits from 
more events relative to the trilepton final states investigated here. 

This analysis is based on the measurement of the $WZ$ production cross section 
in $\ell\nu\ell\ell$ final states~\cite{WZpub} using 4.1~{\rm fb}$^{-1}$ of 
integrated luminosity of $p\bar{p}$ collisions at $\sqrt{s} = 1.96$~{\rm TeV}. 
We extend the selection by analyzing events with any number of jets  in the 
final state and investigate observables that are sensitive to the signal topology 
in order to select events with $WZ \rightarrow \ell\nu\ell\ell$ decays that  
originate from the pair production of top quarks.


\section{Object Reconstruction}
An electron is identified from the properties of clusters of energy deposited in 
the central calorimeters (CC), end cap calorimeters (EC), or intercryostat 
detector (ICD) that match a track reconstructed in the central tracker. 
Because of the lack of far forward coverage of the tracker, we define EC 
electrons only within $1.5 < |\eta| < 2.5$. The calorimeter clusters in the CC 
and EC are required to pass the isolation cut $$\frac{E_{\rm tot} (\Delta {\cal R} < 0.4)  - 
E_{\rm EM} (\Delta {\cal R} < 0.2)}{E_{\rm EM} (\Delta {\cal R} < 0.2)} < 0.1$$
for ``loose" electrons and $< 0.07$ for ``tight" electrons, where $E_{\rm tot}$ 
is the total energy in the EM and hadronic calorimeters, $E_{\rm EM}$ is the 
energy found in the EM calorimeter only, and 
$\Delta {\cal{R}} = \sqrt{(\Delta\phi)^2+ (\Delta\eta)^2}$, where $\phi$ is the 
azimuthal angle. For the intercryostat region (ICR), $1.1 < |\eta| < 1.5$, we form 
clusters from the energy deposits in the CC, ICD, or EC detectors. These clusters 
are identified as electrons if they pass a neural network requirement that is based 
on the characteristics of the shower and associated track information. A muon 
candidate is reconstructed from track segments within the muon system that are 
matched to a track reconstructed in the central tracker. The trajectory of the muon 
candidate must be isolated from other tracks within a cone of $\Delta {\cal {R}} < 0.5$, 
with the sum of the tracks' transverse momenta, $p_{T}$, in a cone less than 
$4.0~{\rm GeV}$ for ``loose" muons and less than $2.5~{\rm GeV}$ for ``tight" 
muons. Tight muon candidates must also have less than $2.5~{\rm GeV}$ of 
calorimeter energy in an annulus of $0.1 < \Delta {\cal {R}} < 0.4$. Jets are 
reconstructed from the energy deposited in the CC and EC calorimeters, using 
the ``Run II midpoint cone'' algorithm~\cite{RunIIcone} of size 
$\Delta {\cal{R}} = 0.5$, within $|\eta| < 2.5$. 

\section{Signal and Background Monte Carlo Simulations}
Monte Carlo (MC) samples of $WZ$ and $ZZ$ background events are produced 
using the {\sc pythia} generator~\cite{pythia}. The production of the $W$ and 
$Z$ bosons in association with jets ($W + $jets, $Z + $jets), collectively referred 
to as $V + $jets, as well as $t\bar{t}$ processes are generated using 
{\sc alpgen}~\cite{alpgen} interfaced with {\sc pythia} for parton evolution and 
hadronization. In all samples the CTEQ6L1 parton distribution function (PDF) 
set is used, along with $m_{\rm t} = 172.5$~{\rm GeV}. The $t\bar{t}$ cross 
section is set to the SM value at this top quark mass, i.e., 
$\sigma_{t\bar{t}} = 7.46 ^{+0.48}_{-0.67}$ {\rm pb}~\cite{ttbar-cross-sec}. 
This uncertainty is mainly due to the scale dependence, PDFs, and the 
experimental uncertainty on $m_{\rm t}$~\cite{top_wa}.

All MC samples are passed through a {\sc geant}~\cite{geant} simulation of 
the D0 detector and overlaid with data events from random beam crossings 
to account for the underlying event. The samples are then corrected for the 
luminosity dependence of the trigger, reconstruction efficiencies in data, and
the beam position. All MC samples are normalized to the luminosity in data 
using NLO calculations of the cross sections, and are subject to the same 
selection criteria as applied to data.

The signal process is generated using the {\sc pythia} generator with the decay 
$t\rightarrow Zq$ added. The $Z$ boson helicity is implemented by reweighting 
an angular distribution of the positively charged lepton in the decay 
$t \to Z q \to \ell^+ \ell^- q$ using {\comphep}~\cite{comphep}, modified by 
the addition of  the Lagrangian of Eq.~\ref{eq:fcnc_lagrangian}. The variable 
$\cos\theta^*$ used for the reweighting is defined by the angle $\theta^*$ 
between the $Z$ boson's momentum in the top quark rest frame and the 
momentum of the positively charged lepton in the $Z$ boson rest frame. We 
assume in the analysis that the vector and axial vector couplings, as introduced 
in Eq.~\ref{eq:fcnc_lagrangian}, are identical to the corresponding couplings for 
neutral currents (NC) in the SM, i.e., $v_{tuZ}=1/2-4/3 \sin^2\theta_W = 0.192$ 
and $a_{tuZ}=1/2$, where $\sin^2\theta_W=0.231$. To study the influence of 
different values of the couplings, we also analyse the following cases: (i.a) 
$v_{tuZ}=1, a_{tuZ}=0$; (i.b) $v_{tuZ}=0, a_{tuZ}=1$; and (ii) 
$v_{tuZ}=a_{tuZ}=1/\sqrt{2}$. As expected, the first two give identical results. 
The difference obtained by using cases (i), (ii), and using the values of the SM 
NC couplings is included as systematic uncertainty. Therefore, our result is 
independent of the actual values of $v_{tqZ}$ and $a_{tqZ}$. Since we do not 
distinguish $c$ and $u$ quark jets our results are valid also for $u$ and $c$ 
quarks separately.

The total selection efficiency, calculated as a function of 
$B =  \Gamma(t\rightarrow Zq)/\Gamma_{\rm tot}$, where $\Gamma_{\rm tot}$ 
contains $t\rightarrow Wb$ and $t\rightarrow Zq$ decays only, can be written as
\begin{equation}
\label{eq:effttbar}
\epsilon_{t\bar{t}} = (1-B)^2 \cdot \epsilon_{t\bar{t} \to W^{+} b W^{-}
  \bar{b}} + 2B(1-B) \cdot \epsilon_{t\bar{t} \to Z q W^{-} \bar{b}}
  + B^2 \cdot \epsilon_{t\bar{t} \to Z q Z \bar{q}},
\end{equation}
where the efficiency $\epsilon_{t\bar{t} \to W^{+} b W^{-} \bar{b}}$ for the SM 
$t\bar{t}$ background contribution is used, along with the efficiencies 
$\epsilon_{t\bar{t} \to Z q W^- \bar{b}}$ and $\epsilon_{t\bar{t} \to Z q Z \bar{q}}$
that include the FCNC top quark decays.

\section{Event Selection and Signal Acceptance}
We consider four independent decay signatures: $eee+\MET + X$,
$ee\mu+\MET + X$, $\mu\mu e+\MET + X$, and $\mu\mu\mu+\MET + X$,
where $X$ is any number of jets. We require the events to have at least three 
lepton candidates with $p_T > 15$~{\rm GeV} that originate from the same 
$p\bar{p}$ interaction vertex and are separated from each other by 
$\Delta {\cal{R}} > 0.5$. Jets are excluded from consideration unless they 
have $p_T>20$~{\rm GeV}. We also require that the jets are separated from 
electrons by $\Delta {\cal{R}} > 0.5$. There is no fixed separation cut between the 
muon and jets but the muon isolation requirement rejects most muons within 
$\Delta {\cal{R}} < 0.4$ of a jet. The event must also have $\MET > 20$~{\rm GeV}, 
which is calculated from the energy found in the calorimeter cells and $p_{T}$ 
corrected for any muons reconstucted in the event. Furthermore, all energy 
corrections applied to electrons and jets are propagated through to the $\MET$. 

Events are selected using triggers based on electrons and muons. There are 
several high-$p_T$ leptons from the decay of the heavy gauge bosons 
providing a total trigger efficiency for all signatures of $98\% \pm 2\%$.

To identify the leptons from the $Z$ boson decay, we consider only pairs of 
electrons or muons, additionally requiring them to have opposite electric charges. 
If no lepton pair is found within the invariant mass intervals of 74--108~{\rm GeV} 
($ee$), 65--115~{\rm GeV} ($\mu\mu$) or 60--120~{\rm GeV} ($ee$, with one electron 
in the ICR) the event is rejected, else, the pair that has an invariant mass closest to 
the $Z$ boson mass $M_{Z}$ is selected as the $Z$ boson. The lepton with the highest 
$p_{T}$ of the remaining muons or CC/EC electrons in the event is selected as 
originating from the $W$ boson decay. From simulation, this assignment of the 
three charged leptons to $Z$ and $W$ bosons is found to be  $\approx$ 100\% 
correct for $ee\mu$ and $\mu\mu e$, and about 92\% and 89\% for the $eee$ 
and $\mu\mu\mu$ channels, respectively. 

Thresholds in the selection criteria are the same as in Ref.~\cite{WZpub} 
and the acceptance multiplied by efficiency results are summarized in 
Table~\ref{Axeff} for the FCNC signal. These values are calculated with respect 
to the total rate expected for all three generations of leptonic $W$ and $Z$ decays.

\begin{table*}[t]
\begin{center}
\caption{Final efficiencies in \% including detector and kinematic acceptance as 
well as detector efficiencies for each decay signature as a function of jet 
multiplicity $n_{\rm jet}$. The efficiency, $\epsilon$, is defined assuming fully 
leptonic decays of the vector bosons from top quarks, as defined as in 
Eq.~\ref{eq:effttbar}. The statistical and systematic uncertainties have been 
added in quadradure.}
\begin{tabular}{|l|c|c|c|c|} \hline
$n_{\rm jet}$ & Inclusive       & $0$                           & $1$             & $\geq 2$        \\ \hline 
Channel       & \multicolumn{4}{c|}{$\epsilon_{t\bar{t} \to ZqW^-\bar{b}}$ (\%)} \\ \hline
$eee$         & $1.65 \pm 0.24$ & $(7.65 \pm 1.45)\cdot10^{-2}$ & $0.57 \pm 0.09$ & $1.00 \pm 0.15$ \\ \hline
$ee\mu$       & $1.92 \pm 0.18$ & $(6.77 \pm 1.05)\cdot10^{-2}$ & $0.58 \pm 0.06$ & $1.17 \pm 0.11$ \\ \hline
$\mu \mu e$   & $1.23 \pm 0.13$ & $(3.37 \pm 0.73)\cdot10^{-2}$ & $0.34 \pm 0.04$ & $0.84 \pm 0.10$ \\ \hline
$\mu \mu \mu$ & $1.48 \pm 0.19$ & $(3.05 \pm 0.74)\cdot10^{-2}$ & $0.38 \pm 0.06$ & $1.06 \pm 0.15$ \\ \hline
Channel       & \multicolumn{4}{c|}{$\epsilon_{t\bar{t} \to ZqZ\bar{q}}$ (\%)} \\ \hline
$eee$         & $1.22 \pm 0.18$ & $(4.69 \pm 0.68)\cdot10^{-2}$ & $0.41 \pm 0.06$ & $0.76 \pm 0.11$ \\ \hline
$ee\mu$       & $3.75 \pm 0.38$ & $(1.07 \pm 0.11)\cdot10^{-1}$ & $1.08 \pm 0.11$ & $2.56 \pm 0.25$ \\ \hline
$\mu \mu e$   & $1.47 \pm 0.16$ & $(3.22 \pm 0.57)\cdot10^{-2}$ & $0.38 \pm 0.05$ & $1.06 \pm 0.32$ \\ \hline
$\mu \mu \mu$ & $2.76 \pm 0.36$ & $(3.63 \pm 0.69)\cdot10^{-2}$ & $0.63 \pm 0.09$ & $2.10 \pm 0.28$ \\ \hline
\end{tabular}
\label{Axeff}
\end{center}
\end{table*}

\section{Data-Driven Backgrounds}
In addition to SM $WZ$ production, the other major background is from processes 
with a $Z$ boson and an additional object misidentified as the lepton from the 
$W$ boson decay (e.g., from $Z + $jets, $ZZ$, and $Z \gamma$). A small 
background contribution is expected from processes such as $W + $jets and 
SM $t\bar{t}$ production. The $WZ$, $ZZ$, and $t\bar{t}$ backgrounds are 
estimated from the simulation, while the $V + $jets and $Z\gamma$ backgrounds 
are estimated using data-driven methods.

One or more jets in $V + $jets events can be misidentified as a lepton from 
$W$ or $Z$ boson decays. To estimate this contribution, we define a {\it false} 
lepton category for electrons and muons. A {\it false} electron is required to 
have most of its energy deposited in the electromagnetic part of the calorimeter 
and satisfy calorimeter isolation criteria for electrons, but have a shower shape 
inconsistent with that of an electron. A muon candidate is categorized as 
{\it false} if it fails isolation criteria, as determined from the total $p_{T}$ of 
tracks located within a cone $\Delta {\cal{R}} = 0.5$ around the muon. These 
requirements ensure that the {\it false} lepton is either a misidentified jet or 
a lepton from the semi-leptonic decay of a heavy-flavor quark. Using a sample 
of data events, collected using jet triggers with no lepton requirement, we measure 
the ratio of misidentified leptons passing two different selection criteria, {\it false} 
lepton and signal lepton, as a function of $p_T$ in three bins, $n_{\rm jet}$ = 0, 1, 
and $\geq 2$, where $n_{\rm jet}$ is the number of jets. We then select a sample 
of $Z$ boson decays with at least one additional {\it false} lepton candidate for 
each final state signature. The contribution from the $V + $jets background is 
estimated by scaling the number of events in this $Z+${\it false} lepton sample 
by the corresponding $p_T$-dependent misidentification ratio.

Initial or final state radiation in $Z\gamma$ events can mimic the signal process 
if the photon either converts into an $e^{+}e^{-}$ pair or is wrongly matched with 
a central track mimicking an electron and the $\MET$ is mismeasured. As a result 
the $Z\gamma$ process is a background to the  final state signatures with 
$W\to e\nu$ decays. To estimate the contribution from this background, we model 
the kinematics of these events using the $Z\gamma$ NLO MC simulation~\cite{Baur}.  
We scale this result by the rate at which a photon is misidentified as an electron. 
This rate is obtained using a data sample of $Z\to\mu\mu$ events containing a 
radiated photon, as it offers an almost background-free source of photons. The 
invariant mass $M(\mu\mu\gamma)$ is reconstructed and required to be consistent 
with the $Z$ boson mass. The $Z\rightarrow \mu\mu$ decay is chosen to avoid any 
ambiguity when assigning the electromagnetic shower to the final state photon candidate. 
As the $Z\gamma$ NLO MC does not model recoil jets, {\sc pythia} MC samples are used to 
estimate $Z\gamma$ background jet multiplicities and $\MET$.  As the {\sc pythia} 
samples do not contain events with final state radiation, we find the fraction of 
$Z \gamma$ events in data and {\sc pythia} MC that pass our $\MET$ cut and take the 
difference as a systematic uncertainty.

\section{Results}
After all selection criteria have been applied, we observe a total of $35$ candidate events 
and expect $31.7 \pm 0.3({\rm stat}) \pm 3.9 ({\rm syst})$ background events from SM 
processes. The statistical uncertainty is due to MC statistics while the sources of systematic 
uncertainties are discussed later. Table~\ref{tab:EvYieldsJetBins} summarizes the number 
of events in each $n_{\rm jet}$ bin. The observed number of candidate and background 
events for each topology, summing over $n_{\rm jet}$, are summarized in 
Table~\ref{tab:EvYieldsInclJets}. In Tables~\ref{tab:EvYieldsJetBins} and~\ref{tab:EvYieldsInclJets} 
and in all the following figures, we assume a $B$ of 5\%.

\begin{table*}[t]
\caption{Number of observed events, expected number of $t\bar{t}$ FCNC events, 
and number of expected background events for each $n_{\rm jet}$ bin 
with statistical and systematic uncertainties. The MC statistical uncertainty 
on the $t\bar{t}$ signal is negligible, and we only present the systematic uncertainties. 
We assume $B = 5\%$.}
\begin{tabular}{|c|c|c|c|} \hline 
$n_{\rm jet}$   & $0$                       & $1$                      & $\geq 2$                  \\ \hline
Background      & $25.66 \pm 0.28 \pm 3.26$ & $5.06 \pm 0.14 \pm 0.56$ & $0.92 \pm 0.08 \pm 0.09$  \\ \hline
$t\bar{t}\rightarrow WbZq$ & $0.20 \pm 0.03$   & $1.80 \pm 0.27$   & $3.87 \pm 0.56$ \\ 
$t\bar{t}\rightarrow ZqZq$ & $0.002 \pm 0.001$ & $0.020 \pm 0.003$ & $0.050 \pm 0.007$ \\ \hline
Observed        & 30                        & 4                        & 1                         \\ \hline
\end{tabular}
\label{tab:EvYieldsJetBins}
\end{table*}

\begin{table*}[t]
\caption{Number of observed events, expected number of $t\bar{t}$ FCNC events, 
and number of expected background events for each final state signature 
with statistical and systematic uncertainties. The MC statistical uncertainty 
on the $t\bar{t}$ signal is negligible, and we only present the systematic 
uncertainties. We assume $B = 5\%$.}
\begin{tabular}{|l|c|c|c|c|} \hline
Source      & $eee$                    & $ee\mu$                  & $e\mu\mu$                & $\mu\mu\mu$               \\ \hline
$WZ$        & $6.64 \pm 0.07 \pm 1.19$ & $7.51 \pm 0.08 \pm 1.11$ & $4.75 \pm 0.06 \pm 0.69$ & $6.10 \pm 0.07 \pm 1.00$ \\ 
$ZZ$        & $0.33 \pm 0.03 \pm 0.06$ & $1.76 \pm 0.07 \pm 0.17$ & $0.46 \pm 0.04 \pm 0.07$ & $1.30 \pm 0.06 \pm 0.21$ \\ 
$V$ + jets  & $0.60 \pm 0.13 \pm 0.11$ & $0.40 \pm 0.18 \pm 0.17$ & $0.48 \pm 0.10 \pm 0.01$ & $0.22 \pm 0.05 \pm 0.03$ \\ 
$Z\gamma$   & $0.18 \pm 0.05 \pm 0.08$ & $< 0.001$                & $0.66 \pm 0.07 \pm 0.38$ & $< 0.001$        \\ 
$t\bar{t}\rightarrow WbWb$  & $0.04 \pm 0.01 \pm 0.01$ & $0.04 \pm 0.01 \pm 0.01$ & $0.05 \pm 0.01 \pm 0.01$ & $0.04 \pm 0.01 \pm 0.01$ \\ \hline
Background  & $7.89 \pm 0.16 \pm 1.20$ & $9.71 \pm 0.21 \pm 1.14$ & $6.40 \pm 0.14 \pm 0.79$ & $7.66 \pm 0.11 \pm 1.02$ \\ \hline
$t\bar{t}\rightarrow WbZq$  & $1.57 \pm 0.22$   & $1.73 \pm 0.17$   & $1.17 \pm 0.13$   & $1.41 \pm 0.18$\\ 
$t\bar{t}\rightarrow ZqZq$  & $0.010 \pm 0.001$ & $0.029 \pm 0.003$ & $0.011 \pm 0.001$ & $0.022 \pm 0.003$\\ \hline
Observed    & 8                        & 13                       & 9                        & 5        \\ \hline
\end{tabular}
\label{tab:EvYieldsInclJets}
\end{table*}

To achieve better separation between signal and background, we analyze the 
$n_{\rm jet}$ and $H_{\rm T}$ distributions (defined as the scalar sum of 
transverse momenta of all leptons, jets, and $\MET$), and the reconstructed 
invariant mass for the products of the decay $t \rightarrow Zq$.

\begin{figure}[ht]
 \begin{center}
   \includegraphics[width=0.45\textwidth]{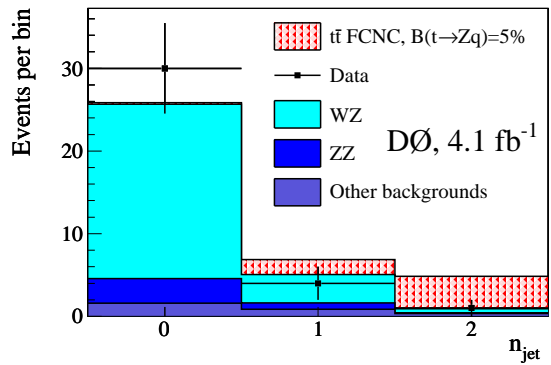}
	\caption{Distribution of $n_{\rm jet}$ for data, for simulated FCNC $t\bar{t}$ signal, 
	and for the expected background. The $ZqZq$ signal is included in the $t\bar{t}$ 
	FCNC contribution but is expected to be small, as can be seen from 
	Tables~\ref{tab:EvYieldsJetBins} and~\ref{tab:EvYieldsInclJets}.}
    \label{fig:jetbins}
  \end{center}
\end{figure}

\begin{figure}[ht]
 \begin{center} 
   \includegraphics[width=0.45\textwidth]{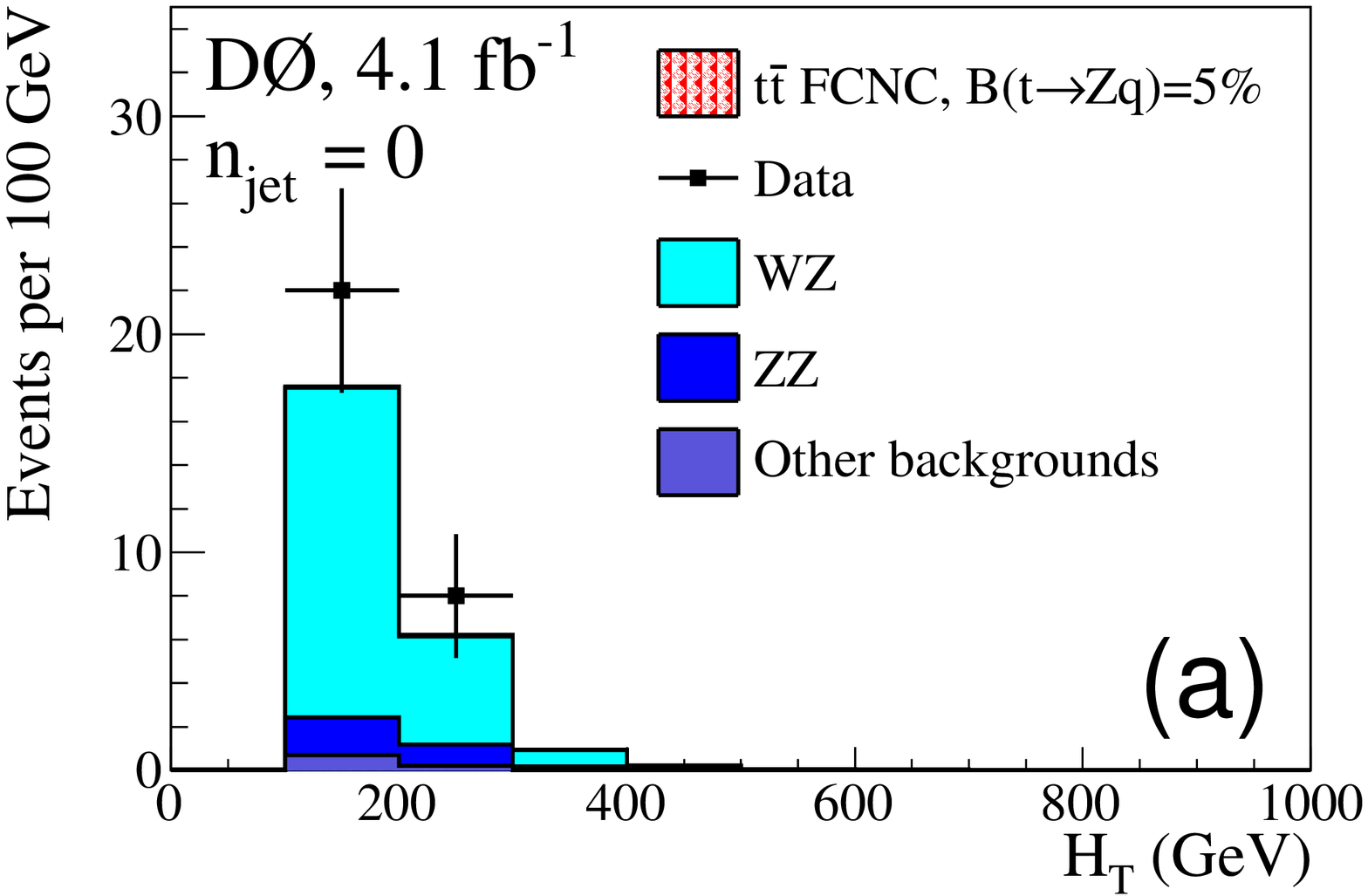}
   \includegraphics[width=0.45\textwidth]{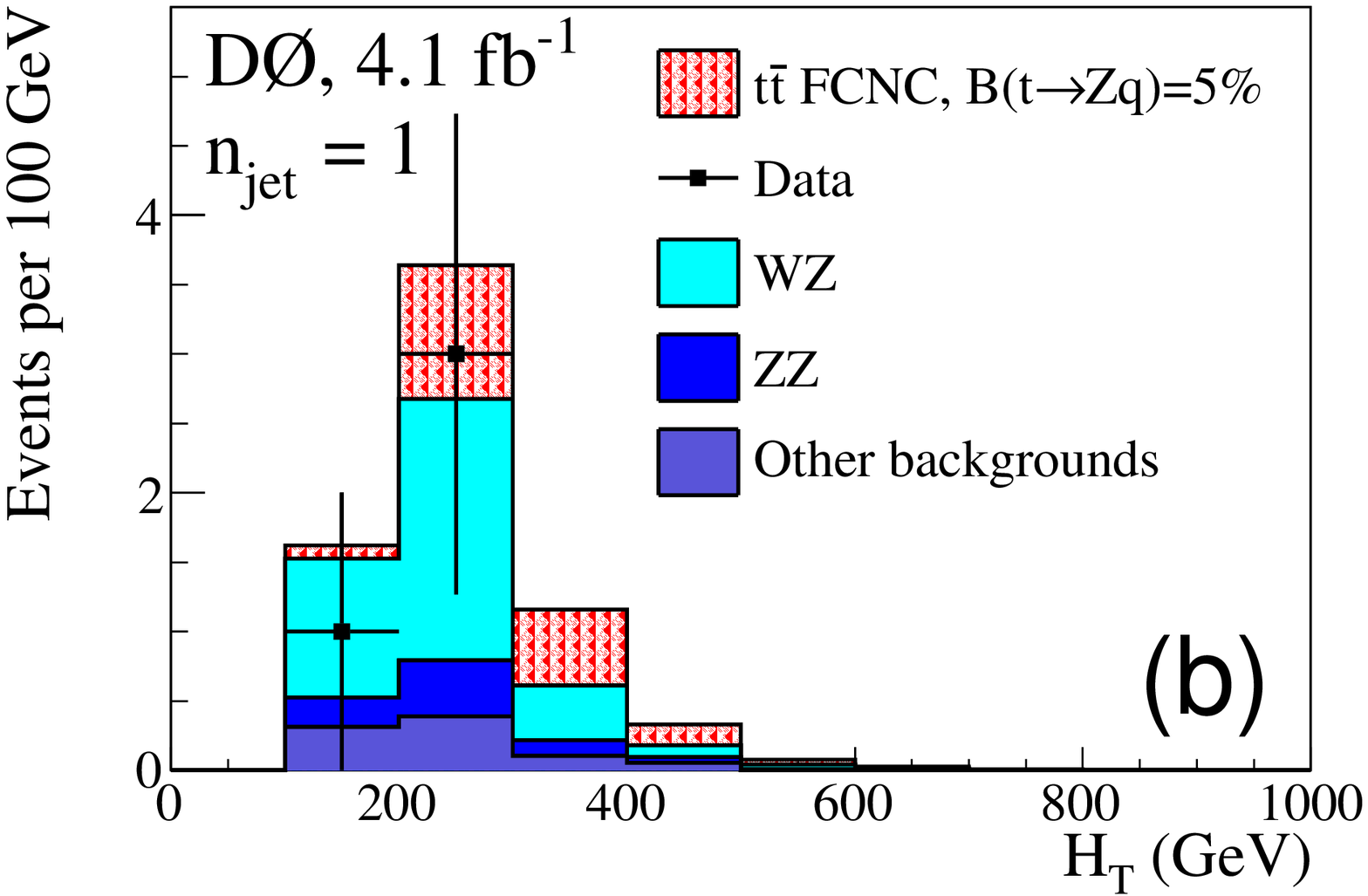}
   \includegraphics[width=0.45\textwidth]{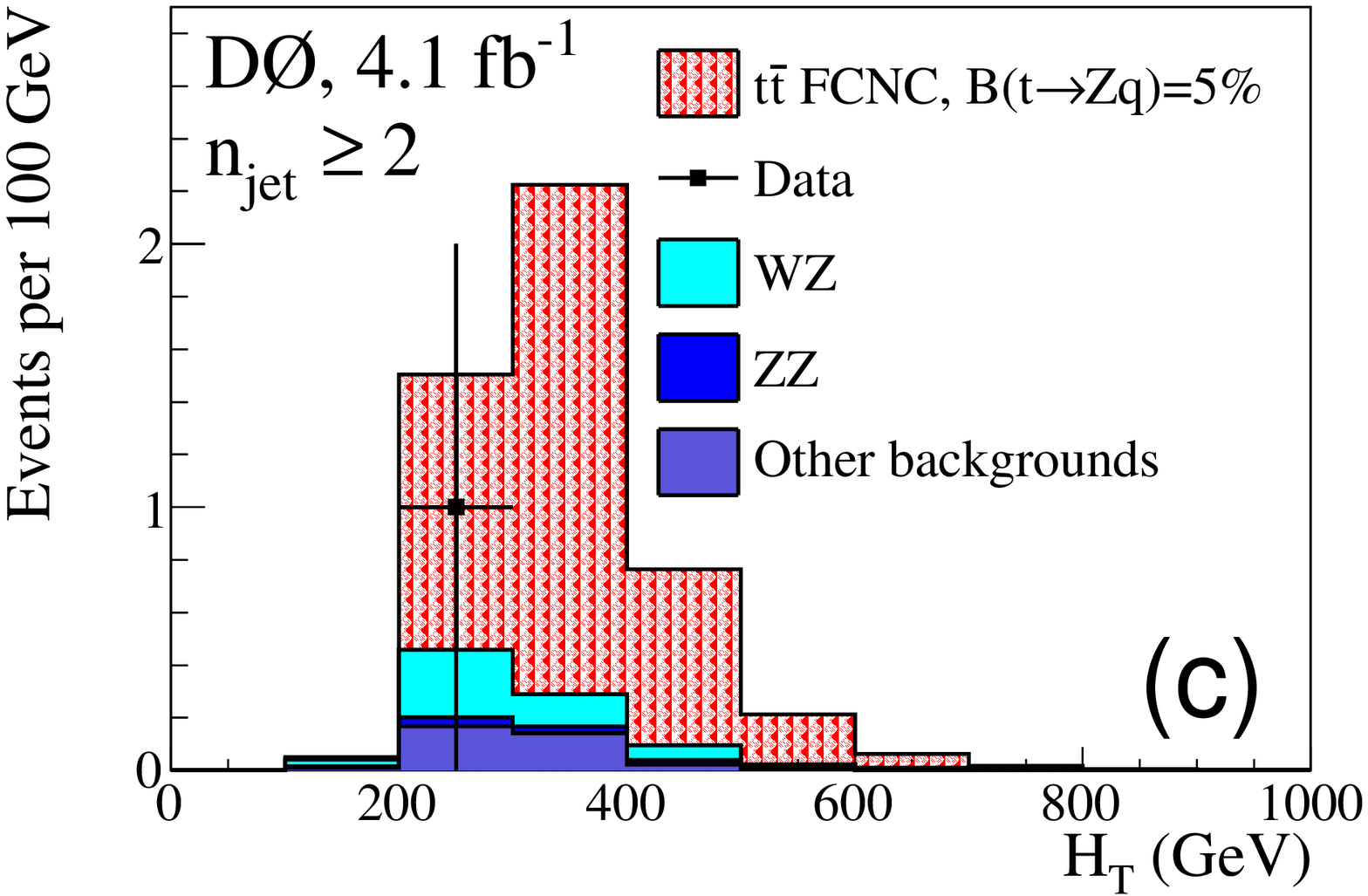}
   \caption{$H_{\rm T}$ distribution of data, FCNC $t\bar{t}$ signal, 
   and expected background for events with (a) $n_{\rm jet} = 0$, 
   (b) $n_{\rm jet} = 1$, and (c) $n_{\rm jet} \geq 2$.}
   \label{fig:ht_all} 
  \end{center} 
\end{figure}

The jet multiplicities in data, SM background, and in FCNC top quark decays 
are shown in Fig.~\ref{fig:jetbins}.  
FCNC $t\bar{t}$ production leads to larger jet multiplicities and also a larger 
$H_{\rm T}$. This is shown in Fig.~\ref{fig:ht_all}.

\begin{figure}[ht]
 \begin{center} 
   \includegraphics[width=0.45\textwidth]{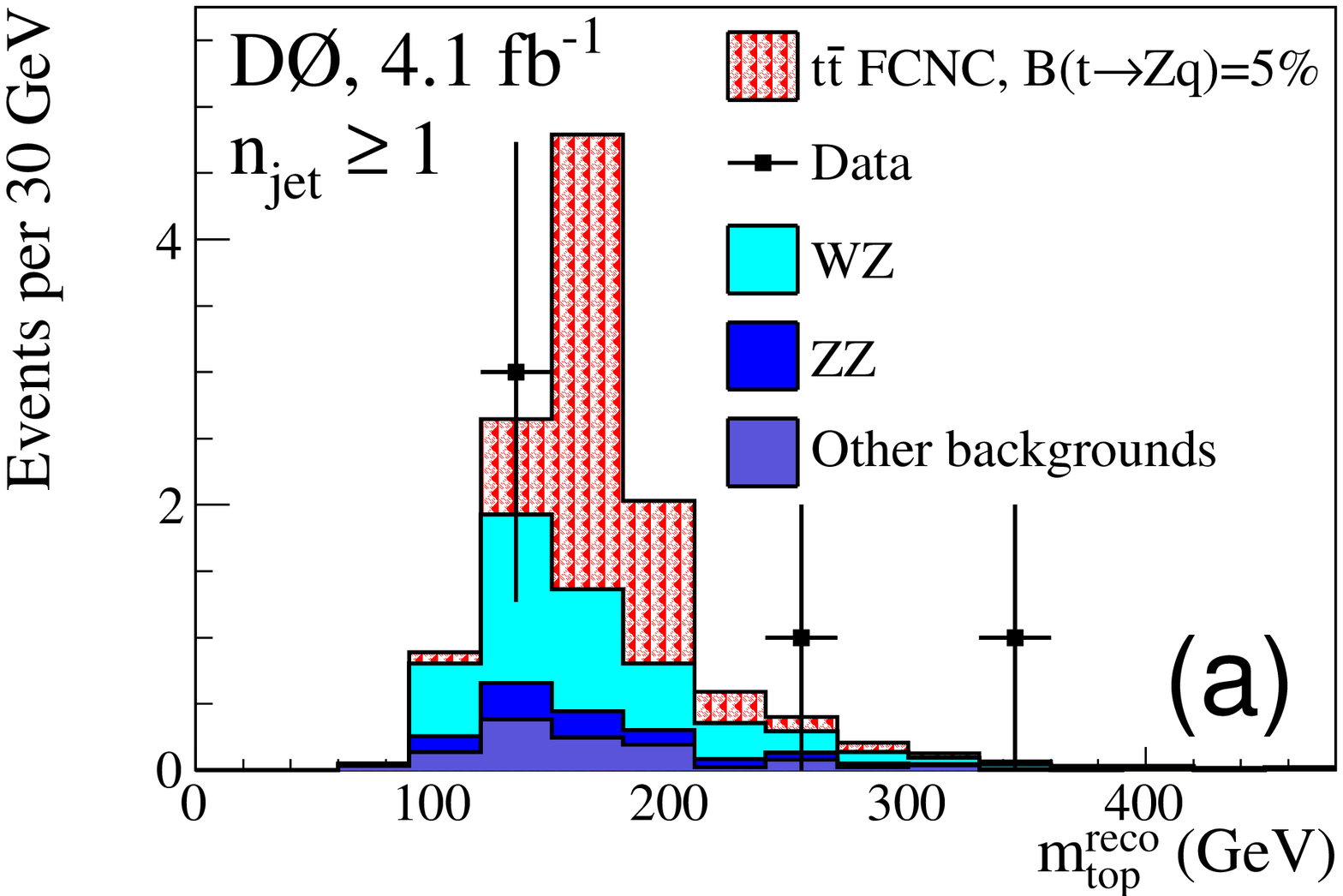}
   \includegraphics[width=0.45\textwidth]{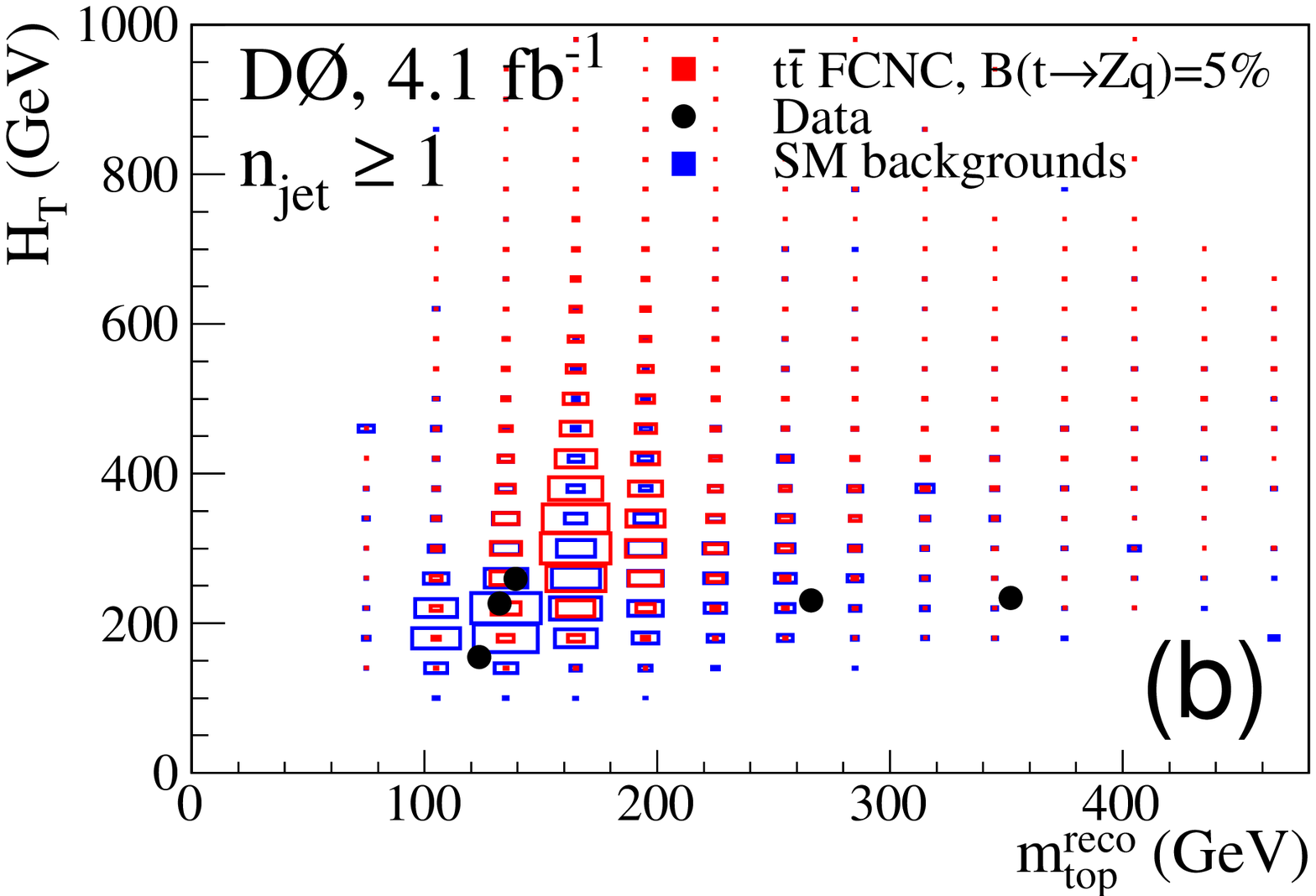}
   \caption{(a) $m_{\rm t}^{\rm reco}$ distribution of data, FCNC $t\bar{t}$ signal, 
   and expected background for events with $n_{\rm jet} \geq 1$; 
   (b) $H_{\rm T}$ vs. $m_{\rm t}^{\rm reco}$ distribution of 
   data, FCNC $t\bar{t}$ signal, and background for events with 
   $n_{\rm jet} \geq 1$. }
   \label{fig:mzjet} 
  \end{center} 
\end{figure}

To further increase our sensitivity we reconstruct the mass of the top quark 
that decays via FCNC to a $Z$ boson and a quark ($t\rightarrow Zq$). In events 
with $n_{\rm jet} = 0$, this variable is not defined. In events with one jet, we 
calculate the invariant mass, $m_{\rm t}^{\rm reco} \equiv M(Z,{\rm jet})$, 
from the 4-momenta of the jet and the identified $Z$ boson, to reconstruct 
$m_{\rm t}$. For events with two or more jets, we use the jet that gives a 
$m_{\rm t}^{\rm reco}$ closest to $m_t=172.5$~{\rm GeV}. 
The $m_{\rm t}^{\rm reco}$ distribution is shown in Fig.~\ref{fig:mzjet}(a). In 
Fig.~\ref{fig:mzjet}(b), we present a 2-dimensional distribution
of $m_{\rm t}^{\rm reco}$ and $H_{\rm T}$.

None of the observables in Figs.~\ref{fig:jetbins} -- \ref{fig:mzjet} show 
evidence for the presence of FCNC in the decay of $t\bar{t}$. We therefore 
set 95\% C.L. limits on the branching ratio $B(t\rightarrow Zq)$. The limits 
are derived from 10 bins of the $H_{\rm T}$ distributions for $n_{\rm jet} = 0, 1$, 
and $\geq 2$. For the channels with $n_{\rm jet} = 1$ and $n_{\rm jet} \geq 2$, 
we split each $H_{\rm T}$ distribution into 4 bins in $m_{\rm t}^{\rm reco}$, 
$m_{\rm t}^{\rm reco} <$ 120~{\rm GeV}, 120 $< m_{\rm t}^{\rm reco} <$ 
150~{\rm GeV}, 150 $< m_{\rm t}^{\rm reco} <$ 200~{\rm GeV}, and 
$m_{\rm t}^{\rm reco} >$ 200~{\rm GeV}. 

\section{Systematic Uncertainties}
When calculating the limit on the branching ratio we consider several sources 
of systematic uncertainty. The systematic uncertainties for lepton-identification 
efficiencies are $15\%$ ($eee$), $11\%$ ($ee\mu$), $9\%$ ($\mu\mu e$), and 
$12\%$ ($\mu\mu\mu$). The systematic uncertainty assigned to the choice of 
PDF is $5\%$. In addition, we assign $9\%$ systematic uncertainty on 
$\sigma_{t\bar{t}}$~\cite{ttbar-cross-sec}. This includes the dependence on
the uncertainty of $m_{\rm t}$~\cite{top_wa}. Furthermore, $m_{\rm t}$ is changed 
from $172.5$~{\rm GeV} to $175$~{\rm GeV} in $t\bar{t}$ MC samples with the 
difference in the result taken as a systematic uncertainty. We vary the $v_{tqZ}$ 
and $a_{tqZ}$ couplings as explained before Eq.~\ref{eq:effttbar}, resulting in a 
1\% systematic uncertainty on the acceptance. Due to the uncertainty on the 
theoretical cross sections for $WZ$ and $ZZ$ production, we assign a 
$10\%$~\cite{vv-cross-sec} systematic uncertainty to each. The major sources 
of systematic uncertainty on the estimated $V + $jets contribution arise from 
the \MET~requirement and the statistics in the multijet sample used to measure 
the lepton-misidentification rates. These effects are estimated independently 
for each signature and found to be between $20\%$ and $30\%$. The systematic 
uncertainty on the $Z\gamma$ background is estimated to be $40\%$ and $58\%$ 
for the $eee$ and $\mu\mu e$ channels, respectively. Uncertainties on jet energy 
scale, jet energy resolution, jet reconstruction, and identification efficiency are 
estimated by varying parameters within their experimental uncertainties. For 
$n_{\rm jet} = 0$ the uncertainty is found to be $1\%$, for $n_{\rm jet} = 1$ it 
is $5\%$, and for $n_{\rm jet} \geq 2$ it is $20\%$. The measured integrated 
luminosity has an uncertainty of $6.1\%$~\cite{lumi}.

\section{Limits Setting}
We use a  modified frequentist approach~\cite{cls} where the signal confidence 
level $CL_s$, defined as the ratio of the confidence level for the signal+background 
hypothesis to the background-only hypothesis ($CL_s = CL_{s+b}/CL_b$), is 
calculated by integrating the distributions of a test statistic over the outcomes
of pseudo-experiments generated according to Poisson statistics for the 
signal+background and background-only hypotheses. The test statistic is 
calculated as a joint log-likelihood ratio (LLR) obtained by summing LLR values 
over the bins of the $H_{\rm T}$ distributions. Systematic uncertainties are 
incorporated via Gaussian smearing of Poisson probabilities for signal and 
backgrounds in the pseudo-experiments. All correlations between signal and 
backgrounds are maintained. To reduce the impact of systematic uncertainties 
on the sensitivity of the analysis, the individual signal and background 
contributions are fitted to the data, by allowing a variation of the background 
(or signal+background) prediction, within its systematic uncertainties~\cite{collie}. 
The likelihood is constructed via a joint Poisson probability over the number of 
bins in the calculation, and is a function of scaling factors for the systematic 
uncertainties, which are given as Gaussian constraints associated with their priors.

\begin{figure}[ht]
\begin{center}
  \includegraphics[width=0.5\textwidth]{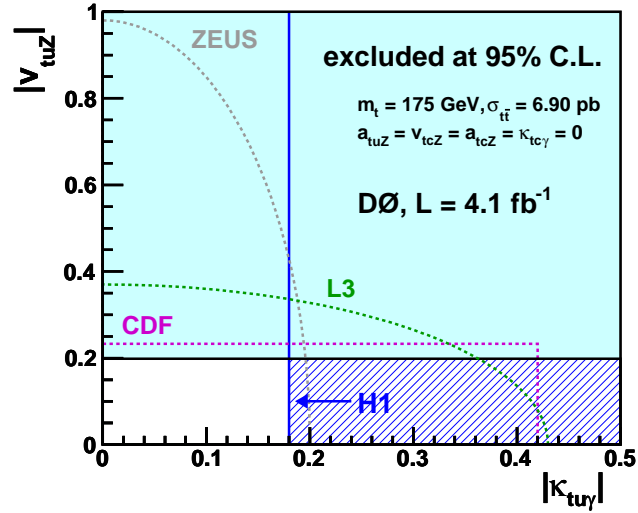}
\end{center}
\caption{Upper limits at the 95\%~C.L. on the anomalous $\kappa_{t u\gamma}$ and 
$v_{tuZ}$ couplings assuming $m_t = 175$~{\rm GeV}. Both D0 and CDF limits on 
$v_{tqZ}$ are scaled to the SM cross section of 
$\sigma_{t\bar{t}} = 6.90$~pb~\cite{ttbar-cross-sec}. Anomalous axial vector 
couplings and couplings of the charm quark are neglected: $a_{tuZ}  =
v_{tcZ} = a_{tcZ} = \kappa_{t c \gamma} = 0$. The scale parameter for
the anomalous dimension-5 coupling $\kappa_{t u \gamma}$ is set to
$\Lambda = m_t = 175$~{\rm GeV}~\cite{fcnc_h1}. Any dependence of the
Tevatron limits on $\kappa_{t u \gamma}$ is not displayed as the change is 
small and at most 6\% for $\kappa_{t u \gamma}=0.5$. The domain excluded 
by D0 is represented by the light (blue) shaded area. The hatched area corresponds 
to the additional domain excluded at HERA by the H1 experiment~\cite{fcnc_h1}. 
Also shown are upper limits obtained at LEP by the L3 experiment~\cite{fcnc_lep} 
(green dashed), at HERA by the ZEUS experiment~\cite{fcnc_zeus} (grey dashed), 
and at the Tevatron by the CDF experiment~\cite{fcnc_tqgamma_cdf,cdflimits} 
(magenta dashed). The region above or to the right of the respective lines is excluded.}
\label{fig:colliders}
\end{figure}

We determine an observed limit of $B(t \rightarrow Zq) < 3.2\%$, with an expected 
limit of $<3.8\%$ at the 95\% C.L. The limits on the branching ratio are converted 
to limits at the 95\%~C.L. on the FCNC vector, $v_{tqZ}$, and axial vector, $a_{tqZ}$,
couplings as defined in Eq.~\ref{eq:fcnc_lagrangian} using the relation given 
in~\cite{fcnc_coupling}. This can be done for any point in the ($v_{tqZ}$, $a_{tqZ}$) 
parameter space and for different quark flavors ($u,c$) since the differences in the 
helicity structure of the couplings are covered as systematic uncertainties in the 
limit on the branching ratio. Assuming only one non-vanishing $v_{tqZ}$ coupling 
($a_{tqZ} = 0$), we derive an observed (expected) limit of $v_{tqZ}<0.19$ ($<0.21$) for 
$m_t=172.5$~{\rm GeV}. Likewise, this limit holds assuming only one non-vanishing 
$a_{tqZ}$ coupling. Figure~\ref{fig:colliders} shows current limits from experiments 
at the LEP, HERA, and Tevatron colliders as a function of the FCNC couplings
$\kappa_{t u \gamma}$ (defined in Ref.~\cite{fcnc_coupling}) and $v_{tuZ}$ for 
$m_t=175$~{\rm GeV}.

\section{Conclusion}
In summary, we have presented a search for top quark decays via FCNC in 
\ttbar\ events leading to final states involving three leptons, an imbalance 
in transverse momentum, and jets. These final states have been explored for 
the first time in the context of FCNC couplings. In the absence of signal, we 
expect a limit of $B(t \rightarrow Zq) < 3.8\%$ and set a limit of 
$B(t \rightarrow Zq) < 3.2\%$ at the $95\%$~C.L. which is currently the 
world's best limit. This translates into an observed limit on the FCNC coupling 
of $v_{tqZ}<0.19$ for $m_t=172.5$~{\rm GeV}. 


\bigskip 

\end{document}